\documentclass[runningheads]{llncs}
\usepackage{multirow}
\usepackage[table,xcdraw]{xcolor}
\usepackage[noadjust]{cite}

\usepackage{amsmath,amssymb,amsfonts}
\usepackage{graphicx}
\usepackage{textcomp}
\usepackage{balance}
\usepackage{multirow}
\usepackage{subfigure}

\usepackage{epsfig,array,epstopdf}
\usepackage{longtable, makecell}
\usepackage{booktabs}
\usepackage{capt-of}
\usepackage{setspace}
\usepackage{color, colortbl}
\usepackage{adjustbox}
\usepackage{mathtools}
\usepackage{algorithm}
\usepackage{algpseudocode}
\usepackage{balance}
\usepackage[T1]{fontenc} 
\usepackage{amsmath}
\newcommand{\repeatthanks}{\textsuperscript{\thefootnote}}

\begin{document}
\title{EEG-based 90-Degree Turn Intention Detection for Brain-Computer Interface}
%
%
\author{Pradyot Anand\inst{1}\thanks{Authors contributed equally to this work.}\orcidID{0009-0009-5304-0253} \and
Anant Jain\inst{2}\repeatthanks\orcidID{0000-0002-7131-8310} \and
Suriya Prakash Muthukrishnan\inst{4}\orcidID{0000-0001-6184-8038} \and
Shubhendu Bhasin\inst{2}\orcidID{0000-0002-8002-9684} \and
Sitikantha Roy\inst{3}\orcidID{0000-0003-2720-4620} \and
Lalan Kumar\inst{1,2}\orcidID{0000-0001-7000-7492}}

\authorrunning{P. Anand et al.}
%
\institute{Bharti School of Telecommunication, Indian Institute of Technology Delhi, India \and Department of Electrical Engineering, Indian Institute of Technology Delhi, India  \and Department of Applied Mechanics, Indian Institute of Technology Delhi, India  \and
Department of Physiology, All India Institute of Medical Sciences, New Delhi, India
\email{pradyotanand282@gmail.com}, 
\email{anantjain@ee.iitd.ac.in},
\email{dr.suriyaprakash@aiims.edu}, 
\email{sbhasin@ee.iitd.ac.in},\\
\email{sroy@am.iitd.ac.in}, 
\email{lkumar@ee.iitd.ac.in}}
\maketitle              
\begin{abstract}
Electroencephalography (EEG)--based turn intention prediction for lower limb movement is important to build an efficient brain-computer interface (BCI) system. This study investigates the feasibility of intention detection of left-turn, right-turn, and straight walk by utilizing EEG signals obtained before the event occurrence. Synchronous data was collected using 31-channel EEG and Inertial Measurement Unit (IMU)-based motion capture systems for nine healthy participants while performing left-turn, right-turn, and straight walk movements. Brain dynamics analysis is presented using EEG source imaging (ESI). The EEG data was preprocessed with steps including Artifact Subspace Reconstruction (ASR), re-referencing, and Independent Component Analysis (ICA) to remove data noise. Feature extraction from the preprocessed EEG data involved computing various statistical measures (mean, median, standard deviation, skew, and kurtosis), and Hjorth parameters (activity, mobility, and complexity). Further, the feature selection was performed using the Random forest algorithm for the dimensionality reduction. The feature set obtained was utilized for 3-class classification using XG boost, gradient boosting, and support vector machine (SVM) with RBF (radial basis function) kernel classifiers in a five-fold cross-validation scheme. Using the proposed intention detection methodology, the SVM classifier using an EEG window of 1.5 s and 0 s time-lag has the best decoding performance with mean accuracy, precision, and recall of 81.23\%, 85.35\%, and 83.92\%, respectively, across the nine participants. The decoding analysis shows the feasibility of turn intention prediction for lower limb movement using the EEG signal before the event onset.

\end{abstract}
\section{Introduction}\label{sec:introduction} 

Brain-computer interface (BCI) enables direct communication between the brain and external devices, bypassing conventional neuromuscular pathways \cite{cavus2023brain}. BCIs measure and interpret brain activations using advanced neuroscience techniques to control external devices and software. Scalp electroencephalography (EEG) is a non-invasive neuroimaging technique that records brain electrical signals due to the activations of neurons. It is popular among non-invasive methods due to its low cost, portability, and high temporal resolution \cite{jain2023embc}. Despite having a lower signal-to-noise ratio than intracortical methods, recent literature shows the feasibility of motor decoding using EEG signals \cite{jain2022premovnet,yen2023recognition,jain2023subject, saini2023bicurnet, blanco2024lower, jain2025esi}. EEG-based BCI systems have been utilized for patients with motor disabilities by enabling the control of external devices such as wheelchairs \cite{al2018review} and robotic arms \cite{ai2023bci}. Moreover, most brain-machine interface (BMI) studies are conducted with stationary participants, as movement can introduce challenging artifacts in the EEG signal \cite{gorjan2022removal}.

Recent studies have shown the feasibility of EEG-based motor intention decoding for lower limb tasks. Decoding of intended user motion using EEG signals is explored for walking-turning and sit-rest-stand tasks \cite{kilicarslan2013high}. Pre-movement EEG signals are utilized to decode sitting and standing intention \cite{bulea2014sitting}. Additionally, EEG-based methodologies are employed to improve obstacle detection during walking \cite{salazar2015analyzing} and recognize gait initiation using supervised classification techniques with score threshold regulation \cite{hasan2019supervised}. The feasibility of predicting gait intentions is demonstrated using pre-movement EEG signals \cite{shafiul2020prediction}. Decoding of self-paced gait intention is reported using EEG before movement initiation \cite{hasan2020asynchronous}.An EEG-based approach is employed to classify sit-to-stand, stand-to-sit, or rest before the event onset \cite{li2023enhanced}. EEG biomarker-based BMI is employed to differentiate between monotonous walking and turn-intentions \cite{quiles2022decoding}.

This study explores EEG-based turn-intention prediction during self-paced lower limb movement. In particular, time-lagged EEG segments before the onset of the events (left-turn, right-turn, and straight-walk) are utilized. First, a dataset consisting of EEG signals is collected from nine participants during the walking-turning task. Subsequently, the EEG data preprocessing, feature extraction, and selection, combined with machine learning models, are performed for turn-intention detection. Various time-lagged EEG segments before event onset are utilized for the classification. The key contributions of this study are: a) Data collection of EEG signals during walking-turning movement. b) exploring Hjorth parameters-based features for detecting self-paced lower-limb movements. c) Investigating 
the effect of various time-lagged EEG segments from the event onset to detect the turn-intention. 

\section{Experiment and Data acquisition}\label{sec:exp_data_acq}
The key objective of the study is to investigate the viability of EEG-based turn intention detection. For this purpose, an experimental setup is designed to record synchronous EEG signals and kinematics data.

\subsection{Participants and Equipment}\label{sec:sub_equip}
The experimental protocol was approved by the Institute Ethics Committee, Indian Institute of Technology Delhi, New Delhi, India, with reference number 2021/P0141. The data was recorded from nine healthy participants (all males, age 27 $\pm$ 3.72 years). EEG data was recorded using 31-channel gel-passive electrodes (BrainCap, Brain Products, Gilching, Germany) with a wireless EEG amplifier (LiveAmp, Brain Products, Gilching, Germany). The EEG electrodes were placed at Fp1, Fp2, F3, F4, C3, C4, P3, P4, O1, O2, F7, F8, T7, T8, P7, P8, Fz, Cz, Pz, FC1, FC2, CP1, CP2, FC5, FC6, CP5, CP6, FT9, FT10, TP9, TP10, according to the International 10-20 systems of EEG electrodes placement. The reference and ground electrodes were placed at FCz and AFz, respectively. The conductive gel was applied to each electrode to ensure good scalp contact and impedance reduction. EEG data was recorded with a sampling frequency of 500 Hz. An IMU-based motion capture system (Xsens Awinda) was utilized to capture the kinematics data. The kinematics data was sampled at the sampling frequency of 100 Hz. The EEG and kinematics data were synchronized using the inbuilt trigger output from the Xsens Awinda station.
\begin{figure*}[t]
\begin{center}	
\includegraphics[scale=0.43]{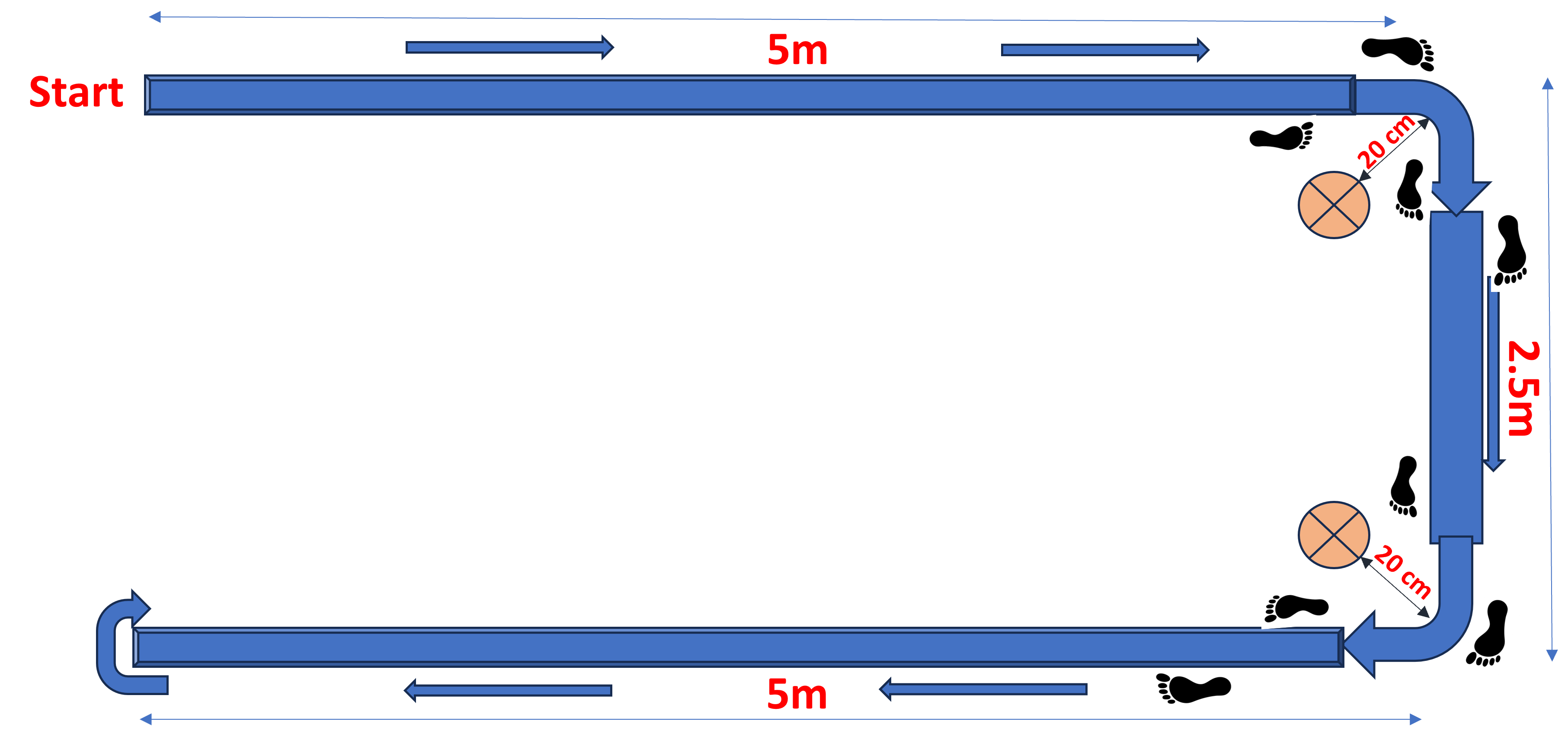}
\caption{Experimental recording paradigm.}
\label{fig:expsetup}
\end{center}
\vspace{-0.5cm}
\end{figure*}
\subsection{Experimental Setup and Paradigm}\label{sec:exp_set_para}
The experimental recording paradigm is shown in Fig. \ref{fig:expsetup}. At the beginning of the experiment, participants stood in a balanced-upright posture. The data collection was performed across five separate runs for each participant. Each run begins with an auditory cue signaling its initialization. Participants began by executing a 90-degree right turn, followed by a 2-3 seconds pause. Subsequently, they prepared for a straight movement trial, halting at the designated endpoint. Following this, participants readied themselves for the second 90-degree turn (a left turn) of the cycle, pausing again for 2-3 seconds after the trial. This sequence was followed by the preparation and execution of the next straight movement trial. This cycle was repeated ten times within each run. An inter-cycle resting period was allowed to avoid physical fatigue. Participants walked and turned at a self-paced speed. Each participant completed a practice run to familiarize themselves with the recording procedure, though this practice run was excluded from the analysis. Data from five runs, each consisting of ten cycles, was recorded for the final analysis. With each cycle having one right turn, one left turn, and two straight walk, each participant performed 50 trials of left and right turns and 100 trials of straight walk.

\begin{figure*}[t]
\begin{center}	
\includegraphics[scale=0.39]{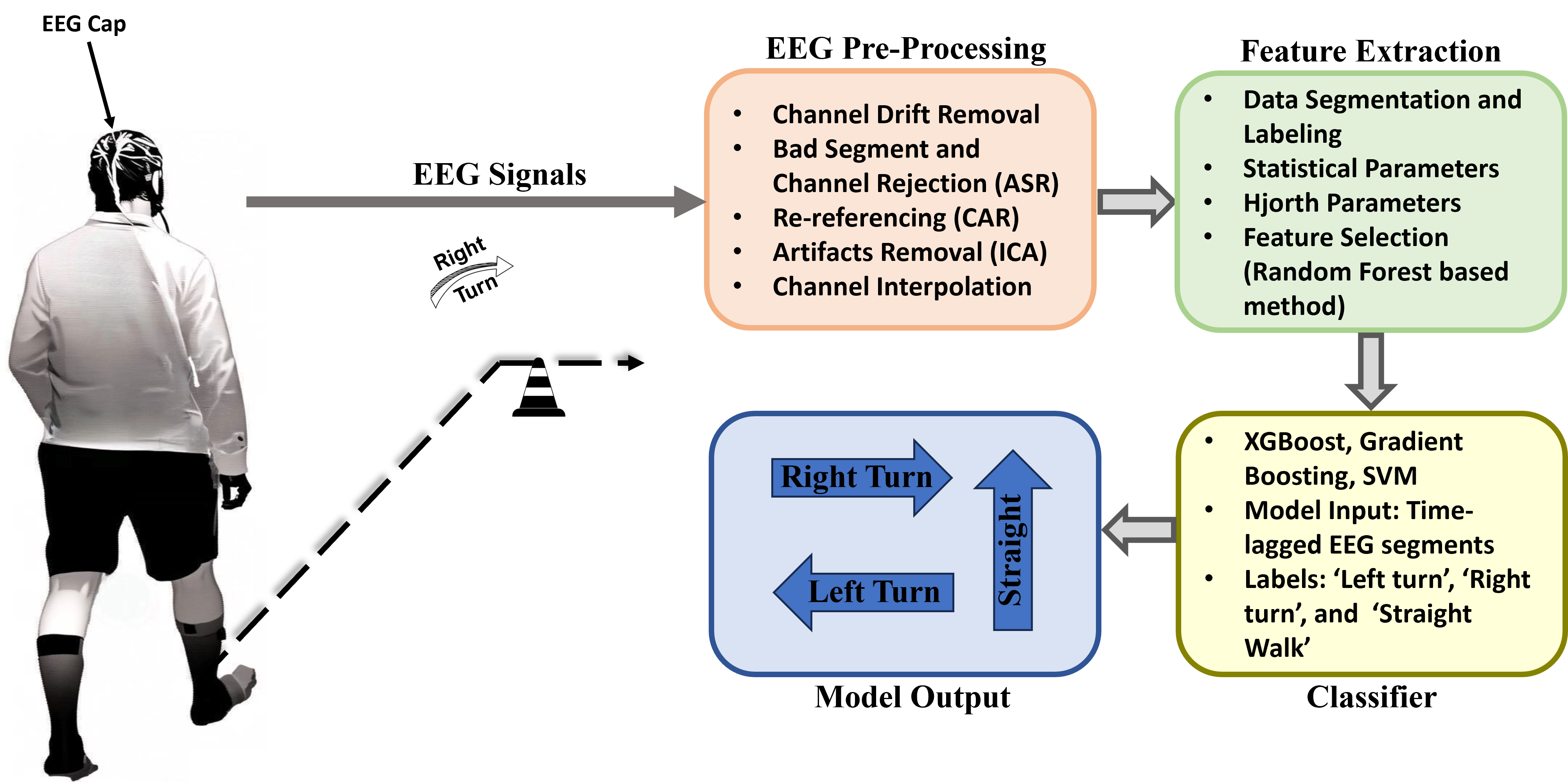}
\caption{Block diagram depicting the proposed framework for turn-intention.}
\label{fig:blockdia}
\end{center}
\vspace{-0.5cm}
\end{figure*}
\section{Methods}\label{sec:methods}
This Section elaborates on the proposed methodology for EEG-based turn-intention detection. The methodology flowchart is illustrated in Fig. \ref{fig:blockdia}. It consists of a detailed description of the EEG preprocessing, data segmentation, feature extraction, and feature selection in Sections \ref{sec:preprocess} $-$ \ref{sec:classification}.
\begin{figure*}[ht!]
    \centering
    \subfigure[]{
        \includegraphics[scale=0.34]{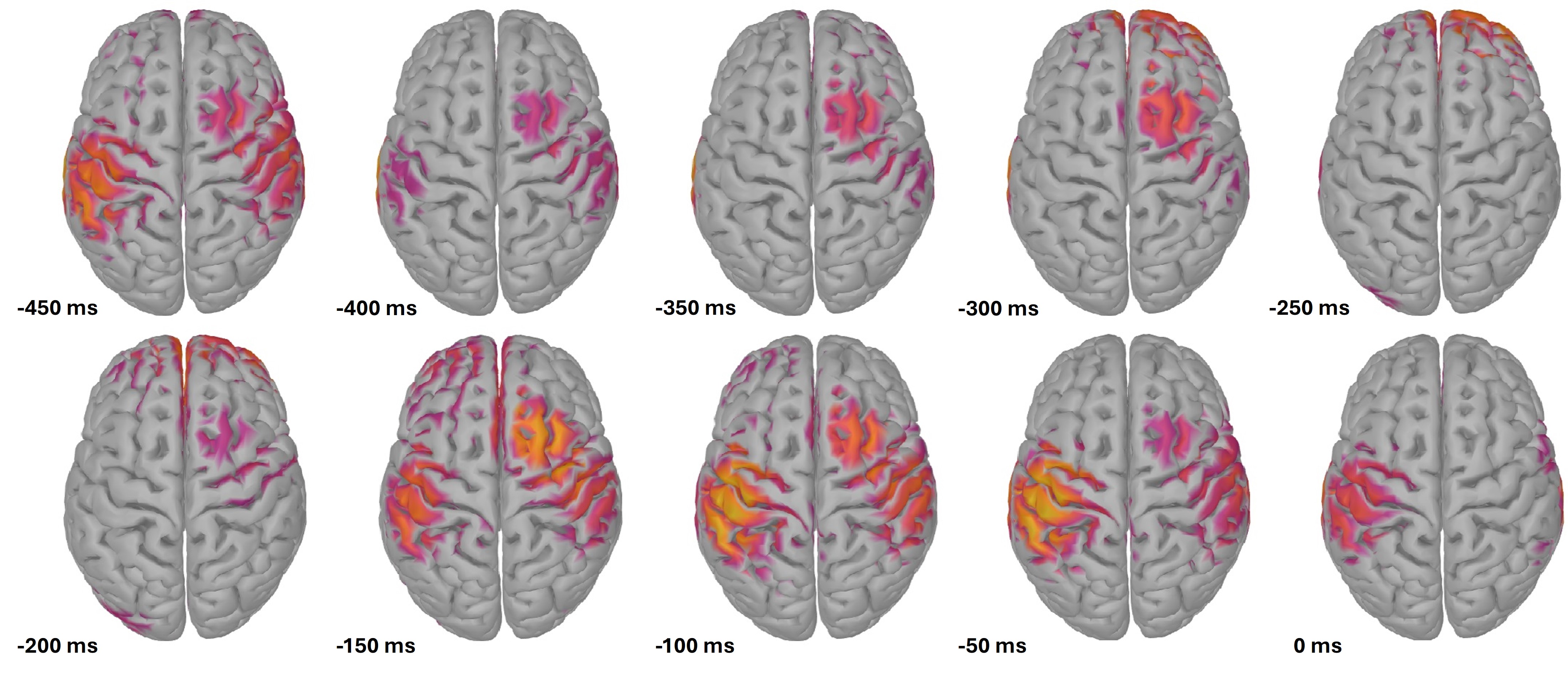}
        \label{fig:figure1}
    }
    \hfill
    \subfigure[]{
        \includegraphics[scale=0.34]{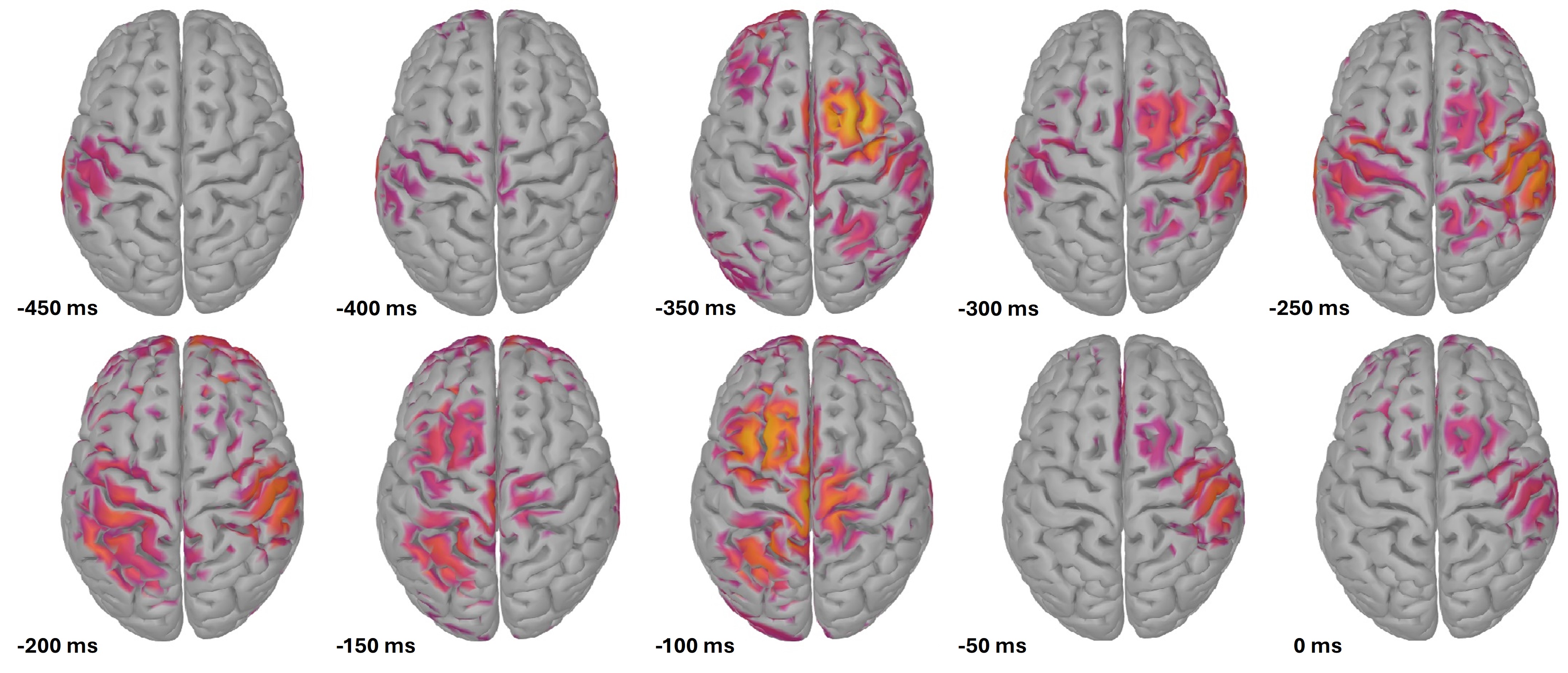}
        \label{fig:figure2}
    }
    \hfill
    \subfigure[]{
        \includegraphics[scale=0.34]{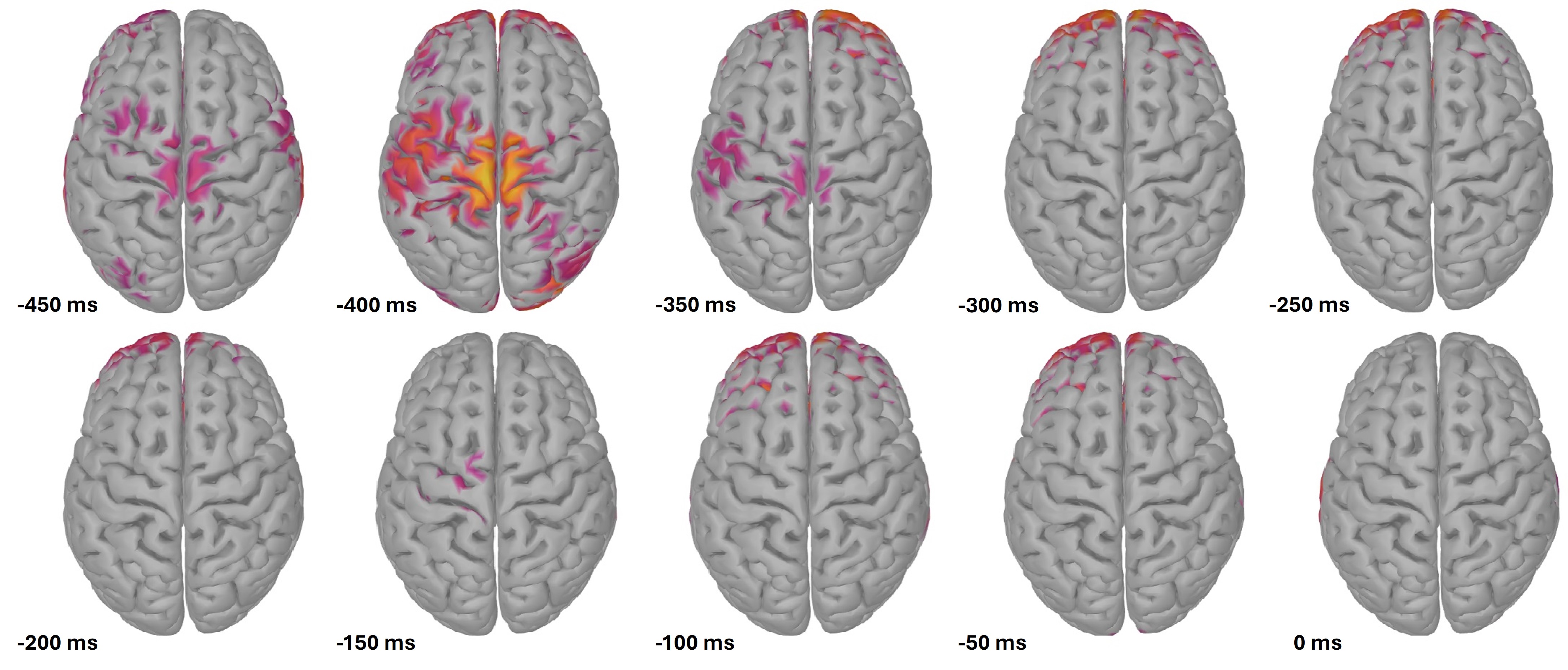}
        \label{fig:figure3}
    }
    \caption{Brain source localization using sLORETA at different time stamps for (a) Left turn, (b) Right turn, and (c) Straight walk.}
    \label{fig:esi}
\end{figure*}
\subsection{EEG Preprocessing}\label{sec:preprocess}

The EEG signal was preprocessed using a MALTAB-based plugin EEGLAB \cite{delorme2004eeglab}. EEG data was high-pass filtered with a cutoff frequency of 0.5 Hz to remove channel drift. The denoising of the recorded EEG signal was performed in two steps: 1) Artifact subspace reconstruction (ASR) and 2) Independent component analysis (ICA). The ASR is utilized to reject bad channels and remove short-term high-amplitude artifacts from continuous data. EEG channel rejection was performed if it exhibited a flat signal for over 5 seconds. The standard deviation (SD) value was set to 20 for repairing artifactual bursts from EEG signals. Following ASR, the EEG data was re-referenced using the common average re-referencing technique. The re-referenced EEG signal was then subjected to ICA for further isolation and removing artifacts-related components associated with eye movements or muscle activity. Initially, 31-channel EEG data was recorded. However, some bad channels were removed while applying the ASR algorithm to EEG data. EEG channel interpolation was performed to reconstruct the rejected EEG channels. The spherical interpolation technique was utilized to estimate missing EEG electrode signals.

\subsection{EEG Source Imaging}\label{sec:esi}
EEG source imaging (ESI) was utilized to observe brain activation dynamics prior to the turns or walking. A boundary element method (BEM)--based forward modeling approach was employed to construct a head model using the ICBM MRI template \cite{mazziotta2001probabilistic}. The standardized low-resolution electromagnetic tomography (sLORETA) \cite{pascual2002standardized} was applied to understand the spatiotemporal dynamics of brain cortical sources. EEG source
localization plots for the cortical grand average across turn onset events are shown in Fig. \ref{fig:esi} in the time range of -450 ms to 0 ms with intervals of 50 ms. Here, t=0 ms represents the onset of the right-turn, left-turn, or straight-walk. It may be noted that the primary regions involved in motor planning and execution include the premotor cortex, supplementary motor area (SMA), and primary motor cortex \cite{muthukumaraswamy2013high}. For the left turn, the activations in this region can be seen approximately 150 ms before the actual turn, as shown in Fig.~\ref{fig:esi}(a). Similar activations can be noticed approximately 350 ms before the right turn, as depicted in Fig.~\ref{fig:esi}(b). In Fig.~\ref{fig:esi}(c), brain activations corresponding to straight-walk intention can be observed 400 ms before the movement onset. EEG brain dynamics exhibit significant overlap in cortical activity within the central motor cortex during the three motor tasks. 
Hence, the identification of the cortical loci for lower limb movements using EEG is challenging due to head motion and difficulty in isolating different types of movement. The observation aligns with findings from previous studies, which include studies on lower-limb motor tasks utilizing fMRI signals during joint movements of the lower limbs \cite{kapreli2007lower} and gait imagery \cite{wang2008cortical}, as well as EEG signals during gait activities \cite{delval2020cortical}. This indicates that the EEG signals alone lack the discriminatory information to distinguish between the three classes. Consequently, the feature extraction process was implemented on the EEG signals, and machine learning-based classifiers were subsequently employed to classify between walking and turn intentions. 

\subsection{Data Preparation}\label{sec:data_seg}
The kinematics data was used to determine the events of turn initiation by visually inspecting the recorded data. These events were marked in the EEG data for turn-intention detection. The aim of the study is to detect turn-intention before the onset. Various time-lagged EEG windows were explored to detect left turn, right turn, and straight movement intentions before the execution. A time lag of up to 1000 ms and window size of up to 2.50 sec were utilized for intention decoding. Subsequently, a three-class classification task was conducted, distinguishing between the intentions of `left-turn,' `right-turn,' and `straight-walk'.

\subsection{Feature extraction}\label{sec:feature_extract}

Feature extraction is a crucial step in EEG signal processing aimed at capturing relevant information from the raw EEG data that can distinguish between different states or conditions, such as different types of movements or cognitive tasks.
Three Hjorth parameters and the statistical measures like mean, median, standard deviation, skew, and kurtosis were computed for signals for the resulting sub-windows for all 31 channels to extract distinguishing features from the non-stationary EEG data. Hjorth parameters quantify the variance, $var$, of the EEG signals \cite{hjorth1970eeg}. It includes activity, mobility, and complexity of the signal defined as follows:

\begin{equation*} {\textit{Activity (A)}} ={var}(x(t)).\end{equation*}

\begin{equation*} {\textit{Mobility (M)}} =\sqrt {\frac {var(x'(t))}{var(x(t))}}.\end{equation*}

\begin{equation*} {\textit{Complexity (C)}} =\frac {Mobility(x'(t))}{Mobility(x(t))},\end{equation*}

where x(t) is the signal, $x'(t)=\frac{dx(t)}{dt}$ is the first derivative of the signal.

A total of 8 features, including 5 statistical and 3 Hjorth features, for each channel were calculated. Hence, a feature vector of size 248 for a total of 31-channels was obtained for each trial.
\begin{table}[!t]
\centering
\caption{Mean accuracy, precision, and recall across the participants using various time-lagged windows for three classifiers.}
\scalebox{0.83}{
\centering
\begin{tabular}{c|c|c|ccccc}
\hline
\multirow{2}{*}{\begin{tabular}[c]{@{}c@{}}Performance\\ metric\end{tabular}} & \multirow{2}{*}{\begin{tabular}[c]{@{}c@{}}Lag\\ window (ms)\end{tabular}} & \multirow{2}{*}{Decoders} & \multicolumn{5}{c}{Window size (s)}                                                                                                                  \\ \cline{4-8} 
                                                                              &                                                                            &                           & \multicolumn{1}{c|}{0.50}    & \multicolumn{1}{c|}{1.00}    & \multicolumn{1}{c|}{1.50}             & \multicolumn{1}{c|}{2.00}             & 2.50    \\ \hline \hline
\multirow{15}{*}{\begin{tabular}[c]{@{}c@{}}Accuracy\\ (\%)\end{tabular}}     & \multirow{3}{*}{0}                                                         & XGB                       & \multicolumn{1}{c|}{75.60} & \multicolumn{1}{c|}{76.21} & \multicolumn{1}{c|}{77.93}          & \multicolumn{1}{c|}{77.57}          & 77.29 \\ \cline{3-8} 
                                                                              &                                                                            & GB                        & \multicolumn{1}{c|}{75.22} & \multicolumn{1}{c|}{75.01} & \multicolumn{1}{c|}{77.68}          & \multicolumn{1}{c|}{77.62}          & 77.08 \\ \cline{3-8} 
                                                                              &                                                                            & SVM                       & \multicolumn{1}{c|}{76.92} & \multicolumn{1}{c|}{79.71} & \multicolumn{1}{c|}{\textbf{81.23}} & \multicolumn{1}{c|}{80.90}          & 80.12 \\ \cline{2-8} 
                                                                              & \multirow{3}{*}{250}                                                       & XGB                       & \multicolumn{1}{c|}{73.45} & \multicolumn{1}{c|}{73.33} & \multicolumn{1}{c|}{74.92}          & \multicolumn{1}{c|}{75.09}          & 74.64 \\ \cline{3-8} 
                                                                              &                                                                            & GB                        & \multicolumn{1}{c|}{72.31} & \multicolumn{1}{c|}{73.60} & \multicolumn{1}{c|}{74.18}          & \multicolumn{1}{c|}{74.20}          & 72.03 \\ \cline{3-8} 
                                                                              &                                                                            & SVM                       & \multicolumn{1}{c|}{72.61} & \multicolumn{1}{c|}{78.89} & \multicolumn{1}{c|}{77.78}          & \multicolumn{1}{c|}{76.62}          & 77.05 \\ \cline{2-8} 
                                                                              & \multirow{3}{*}{500}                                                       & XGB                       & \multicolumn{1}{c|}{69.32} & \multicolumn{1}{c|}{73.41} & \multicolumn{1}{c|}{72.94}          & \multicolumn{1}{c|}{73.62}          & 71.73 \\ \cline{3-8} 
                                                                              &                                                                            & GB                        & \multicolumn{1}{c|}{69.38} & \multicolumn{1}{c|}{72.39} & \multicolumn{1}{c|}{72.22}          & \multicolumn{1}{c|}{72.19}          & 70.98 \\ \cline{3-8} 
                                                                              &                                                                            & SVM                       & \multicolumn{1}{c|}{72.24} & \multicolumn{1}{c|}{75.99} & \multicolumn{1}{c|}{76.43}          & \multicolumn{1}{c|}{75.48}          & 74.36 \\ \cline{2-8} 
                                                                              & \multirow{3}{*}{750}                                                       & XGB                       & \multicolumn{1}{c|}{69.44} & \multicolumn{1}{c|}{69.65} & \multicolumn{1}{c|}{74.66}          & \multicolumn{1}{c|}{72.23}          & 73.42 \\ \cline{3-8} 
                                                                              &                                                                            & GB                        & \multicolumn{1}{c|}{67.16} & \multicolumn{1}{c|}{69.91} & \multicolumn{1}{c|}{71.92}          & \multicolumn{1}{c|}{72.18}          & 72.80 \\ \cline{3-8} 
                                                                              &                                                                            & SVM                       & \multicolumn{1}{c|}{73.74} & \multicolumn{1}{c|}{73.96} & \multicolumn{1}{c|}{76.82}          & \multicolumn{1}{c|}{73.53}          & 75.56 \\ \cline{2-8} 
                                                                              & \multirow{3}{*}{1000}                                                      & XGB                       & \multicolumn{1}{c|}{67.24} & \multicolumn{1}{c|}{67.88} & \multicolumn{1}{c|}{74.02}          & \multicolumn{1}{c|}{71.79}          & 71.20 \\ \cline{3-8} 
                                                                              &                                                                            & GB                        & \multicolumn{1}{c|}{67.73} & \multicolumn{1}{c|}{68.22} & \multicolumn{1}{c|}{72.32}          & \multicolumn{1}{c|}{71.32}          & 71.87 \\ \cline{3-8} 
                                                                              &                                                                            & SVM                       & \multicolumn{1}{c|}{71.10} & \multicolumn{1}{c|}{70.97} & \multicolumn{1}{c|}{74.62}          & \multicolumn{1}{c|}{72.92}          & 74.84 \\ \hline \hline
\multirow{15}{*}{\begin{tabular}[c]{@{}c@{}}Precision\\ (\%)\end{tabular}}    & \multirow{3}{*}{0}                                                         & XGB                       & \multicolumn{1}{c|}{80.23} & \multicolumn{1}{c|}{81.17} & \multicolumn{1}{c|}{82.54}          & \multicolumn{1}{c|}{82.28}          & 82.36 \\ \cline{3-8} 
                                                                              &                                                                            & GB                        & \multicolumn{1}{c|}{80.37} & \multicolumn{1}{c|}{79.80} & \multicolumn{1}{c|}{82.14}          & \multicolumn{1}{c|}{82.44}          & 82.58 \\ \cline{3-8} 
                                                                              &                                                                            & SVM                       & \multicolumn{1}{c|}{82.17} & \multicolumn{1}{c|}{84.22} & \multicolumn{1}{c|}{85.35}          & \multicolumn{1}{c|}{\textbf{85.58}} & 85.33 \\ \cline{2-8} 
                                                                              & \multirow{3}{*}{250}                                                       & XGB                       & \multicolumn{1}{c|}{78.93} & \multicolumn{1}{c|}{78.72} & \multicolumn{1}{c|}{80.45}          & \multicolumn{1}{c|}{80.91}          & 80.34 \\ \cline{3-8} 
                                                                              &                                                                            & GB                        & \multicolumn{1}{c|}{77.84} & \multicolumn{1}{c|}{79.68} & \multicolumn{1}{c|}{80.42}          & \multicolumn{1}{c|}{80.59}          & 78.69 \\ \cline{3-8} 
                                                                              &                                                                            & SVM                       & \multicolumn{1}{c|}{78.57} & \multicolumn{1}{c|}{83.83} & \multicolumn{1}{c|}{83.23}          & \multicolumn{1}{c|}{83.02}          & 83.16 \\ \cline{2-8} 
                                                                              & \multirow{3}{*}{500}                                                       & XGB                       & \multicolumn{1}{c|}{75.47} & \multicolumn{1}{c|}{78.42} & \multicolumn{1}{c|}{78.82}          & \multicolumn{1}{c|}{79.53}          & 79.07 \\ \cline{3-8} 
                                                                              &                                                                            & GB                        & \multicolumn{1}{c|}{75.88} & \multicolumn{1}{c|}{78.51} & \multicolumn{1}{c|}{79.35}          & \multicolumn{1}{c|}{78.40}          & 77.84 \\ \cline{3-8} 
                                                                              &                                                                            & SVM                       & \multicolumn{1}{c|}{78.84} & \multicolumn{1}{c|}{81.70} & \multicolumn{1}{c|}{82.46}          & \multicolumn{1}{c|}{82.01}          & 81.28 \\ \cline{2-8} 
                                                                              & \multirow{3}{*}{750}                                                       & XGB                       & \multicolumn{1}{c|}{75.86} & \multicolumn{1}{c|}{76.04} & \multicolumn{1}{c|}{80.61}          & \multicolumn{1}{c|}{78.32}          & 78.86 \\ \cline{3-8} 
                                                                              &                                                                            & GB                        & \multicolumn{1}{c|}{74.66} & \multicolumn{1}{c|}{76.54} & \multicolumn{1}{c|}{78.32}          & \multicolumn{1}{c|}{78.85}          & 78.02 \\ \cline{3-8} 
                                                                              &                                                                            & SVM                       & \multicolumn{1}{c|}{79.83} & \multicolumn{1}{c|}{80.64} & \multicolumn{1}{c|}{82.10}          & \multicolumn{1}{c|}{80.21}          & 81.90 \\ \cline{2-8} 
                                                                              & \multirow{3}{*}{1000}                                                      & XGB                       & \multicolumn{1}{c|}{73.28} & \multicolumn{1}{c|}{74.92} & \multicolumn{1}{c|}{79.57}          & \multicolumn{1}{c|}{78.32}          & 77.76 \\ \cline{3-8} 
                                                                              &                                                                            & GB                        & \multicolumn{1}{c|}{73.92} & \multicolumn{1}{c|}{75.15} & \multicolumn{1}{c|}{77.67}          & \multicolumn{1}{c|}{77.13}          & 77.11 \\ \cline{3-8} 
                                                                              &                                                                            & SVM                       & \multicolumn{1}{c|}{76.59} & \multicolumn{1}{c|}{77.93} & \multicolumn{1}{c|}{80.70}          & \multicolumn{1}{c|}{79.66}          & 81.01 \\ \hline \hline
\multirow{15}{*}{\begin{tabular}[c]{@{}c@{}}Recall\\ (\%)\end{tabular}}       & \multirow{3}{*}{0}                                                         & XGB                       & \multicolumn{1}{c|}{78.52} & \multicolumn{1}{c|}{80.08} & \multicolumn{1}{c|}{81.31}          & \multicolumn{1}{c|}{80.86}          & 80.75 \\ \cline{3-8} 
                                                                              &                                                                            & GB                        & \multicolumn{1}{c|}{78.30} & \multicolumn{1}{c|}{78.41} & \multicolumn{1}{c|}{81.14}          & \multicolumn{1}{c|}{80.97}          & 80.47 \\ \cline{3-8} 
                                                                              &                                                                            & SVM                       & \multicolumn{1}{c|}{79.92} & \multicolumn{1}{c|}{82.64} & \multicolumn{1}{c|}{\textbf{83.92}} & \multicolumn{1}{c|}{83.87}          & 83.37 \\ \cline{2-8} 
                                                                              & \multirow{3}{*}{250}                                                       & XGB                       & \multicolumn{1}{c|}{77.36} & \multicolumn{1}{c|}{77.74} & \multicolumn{1}{c|}{79.20}          & \multicolumn{1}{c|}{78.91}          & 78.75 \\ \cline{3-8} 
                                                                              &                                                                            & GB                        & \multicolumn{1}{c|}{76.30} & \multicolumn{1}{c|}{78.08} & \multicolumn{1}{c|}{78.53}          & \multicolumn{1}{c|}{78.75}          & 77.02 \\ \cline{3-8} 
                                                                              &                                                                            & SVM                       & \multicolumn{1}{c|}{76.86} & \multicolumn{1}{c|}{82.26} & \multicolumn{1}{c|}{81.87}          & \multicolumn{1}{c|}{81.26}          & 81.09 \\ \cline{2-8} 
                                                                              & \multirow{3}{*}{500}                                                       & XGB                       & \multicolumn{1}{c|}{74.41} & \multicolumn{1}{c|}{77.64} & \multicolumn{1}{c|}{77.19}          & \multicolumn{1}{c|}{77.46}          & 76.85 \\ \cline{3-8} 
                                                                              &                                                                            & GB                        & \multicolumn{1}{c|}{75.01} & \multicolumn{1}{c|}{77.36} & \multicolumn{1}{c|}{77.36}          & \multicolumn{1}{c|}{75.85}          & 76.07 \\ \cline{3-8} 
                                                                              &                                                                            & SVM                       & \multicolumn{1}{c|}{77.08} & \multicolumn{1}{c|}{80.31} & \multicolumn{1}{c|}{80.59}          & \multicolumn{1}{c|}{80.03}          & 79.41 \\ \cline{2-8} 
                                                                              & \multirow{3}{*}{750}                                                       & XGB                       & \multicolumn{1}{c|}{75.13} & \multicolumn{1}{c|}{74.96} & \multicolumn{1}{c|}{78.91}          & \multicolumn{1}{c|}{76.18}          & 77.24 \\ \cline{3-8} 
                                                                              &                                                                            & GB                        & \multicolumn{1}{c|}{73.24} & \multicolumn{1}{c|}{74.63} & \multicolumn{1}{c|}{76.40}          & \multicolumn{1}{c|}{76.01}          & 76.23 \\ \cline{3-8} 
                                                                              &                                                                            & SVM                       & \multicolumn{1}{c|}{78.20} & \multicolumn{1}{c|}{79.08} & \multicolumn{1}{c|}{80.58}          & \multicolumn{1}{c|}{77.80}          & 79.35 \\ \cline{2-8} 
                                                                              & \multirow{3}{*}{1000}                                                      & XGB                       & \multicolumn{1}{c|}{72.51} & \multicolumn{1}{c|}{73.07} & \multicolumn{1}{c|}{77.85}          & \multicolumn{1}{c|}{76.18}          & 76.19 \\ \cline{3-8} 
                                                                              &                                                                            & GB                        & \multicolumn{1}{c|}{72.41} & \multicolumn{1}{c|}{73.06} & \multicolumn{1}{c|}{76.67}          & \multicolumn{1}{c|}{74.96}          & 75.51 \\ \cline{3-8} 
                                                                              &                                                                            & SVM                       & \multicolumn{1}{c|}{75.46} & \multicolumn{1}{c|}{75.97} & \multicolumn{1}{c|}{78.80}          & \multicolumn{1}{c|}{77.68}          & 78.69 \\ \hline
\end{tabular}
}
\label{tab:intrasub_3class}
\vspace{0.05cm}
\scriptsize{\\
XGB - XGBoost, GB - Gradient Boosting, SVM - Support Vector Machine}
\end{table}
\subsection{Feature Selection}\label{sec:feature_selection}

Given the high dimensionality of the feature set relative to the sample size, there is a significant likelihood of encountering redundant and noisy features. Therefore, implementing a feature selection method is crucial to eliminate these redundancies. In this study, a Random Forest-based feature selection technique was employed. This method utilizes the feature importance scores generated by the Random Forest model to identify and retain the most relevant features. Applying this procedure resulted in a reduced and more distinct set of features, enhancing the effectiveness of classification. Post-feature selection, the number of features was reduced to 50 across all subjects and classes.

\subsection{Classification}\label{sec:classification}

The primary objective of this study is to develop a practical methodology for real-time BCI systems. In this study, Support Vector Machine (SVM) with RBF kernel, Gradient Boosting, and XGBoost classifiers were employed to address three-class classification. For each subject, the classifiers were trained to distinguish between right turn, left turn, and straight movement intentions. The performance of these classifiers was assessed using five-fold cross-validation. In this procedure, one fold served as the test set, while the remaining folds were utilized to train the classifier. Each fold was used exactly once as the test set, and the performance metrics were averaged across all folds to ensure robustness and reliability.

\section{Results and Discussion}\label{sec:results&discuss}

\subsection{Experimental details}

Various time-lagged EEG windows were chosen corresponding to the left, right, and straight movements onset for intention detection. EEG window sizes varying from 0.5 to 2.5 seconds were explored for turn-intention detection, as shown in Table \ref{tab:intrasub_3class}. The numbers represent the starting and ending times of the window corresponding to the event onset time. For example, the [-2.5, -0.5] window represents a 2-second window with a time lag of 0.5 sec from the event onset. After that, three-class classification was addressed: `left-turn,' `right-turn,' and `straight-walk.' Data segmentation was performed for nine participants corresponding to the respective movements.

The performance metrics utilized in this study were accuracy, precision, and recall. Accuracy represents the number of correct predictions over the total number of predictions. Precision gives the proportion of true positives to the amount of total predicted positives. Recall, also known as sensitivity or true positive rate, measures the proportion of true positive predictions among correct model predictions. These parameters are defined in the following equations:

\begin{equation*}
Accuracy{\text{ }}(\%) =\frac {TP+TN}{TP+TN+FP+FN} \times 100
\end{equation*}

\begin{equation*}
Precision{\text{ }}(\%) =\frac {TP}{TP+FP} \times 100
\end{equation*}

\begin{equation*}
Recall{\text{ }}(\%) =\frac {TP}{TP+FN} \times 100
\end{equation*}


TP, TN, FP, and FN denote true-positive, true-negative, false-positive, and false-negative predictions, respectively. The mean performance metrics across nine participants are presented in Table \ref{tab:intrasub_3class}.

\subsection{Decoder performance}

Table \ref{tab:intrasub_3class} shows the mean classification accuracy, sensitivity, and specificity across the subjects. Various time-lagged EEG windows were utilized for detecting `left-turn,' `right-turn', and `straight-walk' intentions. Three classifiers, XGBoost, gradient boosting, and SVM classifiers, were explored for this purpose. The SVM classifier had the best classification performance among the three classifiers. The SVM model with [-1.5s, 0] window has the highest mean accuracy, and recall of 81.23\%, and 83.92\%, respectively. In terms of precision, the highest value was obtained using the [-2s, 0] EEG window with the SVM model. The SVM model has the best classification performance for other non-zero time-lagged windows as well. For window [-1.25s, -0.25s], it has the highest mean accuracy, precision, and recall of 78.89\%, 83.83\%, and 82.26\%, respectively. This shows the feasibility of turns and walk intention detection 250 msec before the event onset using the EEG signals.

\subsection{EEG time lag analysis}

EEG time lag analysis is performed to investigate the effect of EEG time lag and window size for turn-intention detection. Various EEG time-lag and window sizes, in the range of 0--1000 ms and 0.5--2.5 sec, respectively, are explored. Table \ref{tab:intrasub_3class} presents the mean accuracy values across the participants. It can be noted that the classification performance is prominently decreasing with the increase in time lag. The SVM model with a 0.5-sec window size has the overall lowest decoding performance with the lowest mean accuracy, precision, and recall. While the intention decoding performance of the model with a 1.5 sec window size was the highest.

\subsection{Future work}

The primary focus of this study is to detect movement intentions for 90-degree left and right turns and straight walking. However, this methodology could be used for more complex types of movements, such as U-turns, avoiding obstacles, or following multiple steps in a planned path, the movements that are common in real-life walking. To do this, the experiment setup would include the additional types of movements and mark their starting points using motion data. Future work could also explore combining EEG signals with other data modalities, like muscle activity (EMG) or eye movements, to make accurate intention detection.

\section{Conclusion}\label{sec:conclusion}
This study explores EEG-based turn-intention detection for self-paced movements. EEG data was collected with left-turn, right-turn, and straight walking movements. The event onsets are marked using the simultaneously collected kinematics data. A comprehensive analysis of EEG brain dynamics is performed, and the observed activation patterns are compared with the reported patterns in previous studies on lower-limb motor tasks. Various time-lagged EEG windows before the event onset are investigated for intention detection. Various machine learning models, including XGBoost, Gradient Boosting, and SVM, are utilized to classify the left turn, right turn, and straight walk intention detection. The SVM model with [-1.5s, 0] window has the best overall decoding performance. However, the classification results using [-1.25s, -0.25s] show the feasibility of intention detection before the event onset.

\section*{Acknowledgment}
This research work was supported in part by the DRDO - JATC project with project number RP04191G.

%
%
%
\bibliographystyle{splncs04}

\bibliography{TurnIntNet}
\end{document}